\documentclass{svmult} 

\usepackage{graphicx}        
\usepackage[bottom]{footmisc}

\begin{document}

\title*{An elementary introduction to the Holographic 
Principle
}
\author{Antonio Dobado}
\institute{Departamento de F\'\i sica Te\'orica \\ 
Universidad Complutense de Madrid \\
28040-Madrid, Spain \\ \texttt{dobado@fis.ucm.es}}

\maketitle

Nowadays it is clear that Quantum Mechanics and General Relativity
have been the two most important paradigms in Fundamental Physics
during the last century. The  year 2005 was chosen by the UNESCO
as the World Year of Physics with the aim to commemorate the
Einstein's miraculous year ({\it annus mirabilis}) 1905, where he
shocked the scientific community by presenting in the volume 17 of
Annalen der Physik a set of articles \cite{Einstein1} about the
Photoelectric Effect, the Brownian Motion and Special Relativity.
In the first one Einstein set the concept of light quanta ({\it
Lichtquanten}), which was one of the conceptual cornerstones of
Quantum Mechanics, and in the last two  he set the basis of the
Theory of Relativity that would give rise to General Relativity
several years later. Professor Galindo, to whom this article is
dedicated, has devoted a great deal of his fruitful scientific
career to Quantum Mechanics and General Relativity as a scholar,
researcher and teacher, and from him several generations of
Spanish theoretical physicists benefited. It is really a honor and
a pleasure to have the opportunity to celebrate, almost
simultaneously, the 70th birthday of Alberto Galindo and the
beginning of the World Year of Physics 2005. We will do that here
by reviewing briefly the genesis and present status of the so
called {\it Holographic Principle}. This is because according to
the opinion of many people it could bring some light on the always
obscure connection between Quantum Mechanics and General
Relativity, the big themes of contemporary Fundamental Physics and
the academic life of Alberto Galindo.

\section{The Generalized Second Law}
\label{sec:1 }

One of the most intriguing predictions of General Relativity (GR)
is the existence of horizons in many of its solutions. These
horizons establish limits and boundaries between  different sets
of events determining the causal structure of space-time. Probably
the most popular solution of the GR Einstein's 1915 field
equations \cite{Einstein2} is the Schwarzschild one, which is
believed to represent the external metric of  black-holes  (BH).
These bizarre objects were already predicted to exist at the Age
of Enlightenment ( \cite{Michell} and \cite{Laplace} ), but only
after the advent of GR they have a sound theoretical support (the
term BH was coined by Wheeler in 1967). The Schwarzschild metric
can be generalized to the Kerr-Newman (KN) solution \cite{KN}
which describes the most general stationary BH exterior and in
general has a quite rich structure of horizons and even
connections with multiple universes (see for instance \cite{MTW}
for a detailed description). The only parameters appearing in this
solution are the mass $M$, the angular momentum $J$ and the
electric charge $Q$ of the BH. This surprising result is known in
the literature as the {\it no-hair theorem}. The proof for four
dimensions can be found for instance in \cite{HandE} or
\cite{Wald} (the theorem does not apply to arbitrary number of
dimensions). It is quite paradoxical since it means that the final
state of any system collapsing into a BH, no matter how complex
the system could be, will be described by just three parameters.

In any case the area $A$ of the event BH horizon
  is given in Planck units by:
\begin{equation}
A=4\pi(R_+^2+a^2),
\end{equation}
where
\begin{equation}
R_+=M\left( 1+ \sqrt{1-\frac{Q^2+a^2}{M^2}}      \right)
\end{equation}
and $a=J/M$  (we are assuming $Q^2+a^2\le M^2$). On the horizon
the angular frequency, the surface gravity and the electric
potential are given respectively by: $\Omega= 4\pi a /A$,
$\kappa=4\pi(R_+-M)/A$ and $\Phi=4\pi R_+ Q/A $.

 At the beginning of the seventies, the work of several people
made possible to establish  a set of rules concerning the state
and evolution of classical BH's \cite{BHproperties}. These rules
can
be stated as follows (see for instance \cite{Galindo1}):\\

{\it Rule 0:} If the BH is stationary the surface gravity $\kappa$
on the horizon is constant.\\

{\it Rule 1:} Under small variations of the parameters of a
stationary BH we have:
\begin{equation}
d M = \frac{\kappa}{8\pi} d A +  \Omega d J+\Phi d Q
\end{equation}

{\it Rule 2:} Given a system with an arbitrary number of BH's with
area $A_n$, the temporal evolution is such that the variation of
total horizon area $A$  does not decreases with time:
\begin{equation}
d A = \sum_n A_n \ge 0
\end{equation}\\

{\it Rule 3:} A stationary BH with $\kappa = 0$ is not accessible
by a finite number of steps.\\

These rules lead to Bekenstein to suggest an analogy between {\it
classical} BH and thermodynamics \cite{Bekenstein1}. According to
it a stationary BH behaves like a system in thermodynamic
equilibrium so that the rules above correspond with the laws of
thermodynamics. One important consequence of this analogy is that
the BH has an entropy proportional to its horizon area, i. e.
$S_{BH}= C A$. In addition Bekenstein also enunciated the so
called {\it Generalized Second Law} (GSL). According to it
whenever a system may suffer a gravitational
 collapse the total entropy $S$ must be defined as the sum of the
 standard matter-entropy $S_m$ plus the entropy of the BH, that
 could eventually appear, as defined above, i.e. $S=S_m+S_{BH}$.
 Thus the GSL states that it is this total entropy the one that is
 never decreasing when strong gravitational effects must be
 taken into account. The GSL, by assigning some entropy to the collapsed
 matter inside the BH, obviously solves the {\it no-hair theorem}
 paradox. However this solution is not complete sice the nature of
 the $\exp S_{BH}$ states given rise to this BH entropy remains
 absolutely unknown.

\section{The Information Paradox}
\label{sec:2}

The BH thermodynamic description given by Bekenstein was suddenly
supported in an impressive way through a semi-classical
computation  done by Hawking that showed that BH's do in fact
radiate \cite{Hawking1}. The radiation is thermal corresponding to
a temperature $T=\kappa/ 2\pi$ which set the constant $C$ to one
quarter so that the BH entropy is given by:
\begin{equation}
S_{BH}=\frac{A}{4}
\end{equation}
One of the most important consequences of Hawking radiation is
that BH's lose mass, shrink their surface and eventually disappear
into a cloud of thermal energy. It can be shown that during the
process of gravitational collapse of some system into a BH and its
ulterior decay into radiation the GSL applies, i.e. the total
entropy never decreases (in fact it increases). However BH
evaporation poses a new challenge to our scarce knowledge of the
quantum aspects of gravity. The problem was outlined by Hawking as
a lost of unitarity in the evolution of the collapsed system ({\it
 information paradox})\cite{Hawking2}. It may be understood in a
very simple way considering the case in which the initial state of
the collapsed matter is a pure quantum state. After the Hawking
evaporation of the BH we are left just with thermal radiation
which is not a pure quantum state and must be described by a
density matrix. Therefore the whole evolution of the system cannot
be unitary.

This statement gave rise to a great controversy in the physics
community during the last three decades. Defenders of orthodox
quantum theory such as Coleman, Thorne, Preskill and others have
proposed different mechanisms to scape from any violation of
unitarity. For example it has been argued that, since the Hawking
computation is only approximate (semi-classical), subtle
correlations in the radiation not taken into account, could
maintain the information content of the system. This is for
instance what happens if we burn a volume of the  {\it Encyclop\ae
dia Britannica}. Correlations in the relative motion of the
molecules in the produced smoke encode all the information
contained in the volume. Needless to say that for all practical
purposes the information is completely lost but still the
evolution of the system is unitary.

Another solution considered  for the {\it information paradox} is
that, as far as the final moments of the life of the BH are
determined for some unknown quantum gravity theory, it is possible
that after Hawking evaporation always a remnant is left which
stores the information contained in the BH. However it is
difficult to see how a Plank-scale sized object could carry all
the information of, for example, a collapsed star.

There are more creative solutions like the supposition that the BH
give rise to a new universe to which the information flows. In
this way unitarity is preserved in the whole set of universes but
it is apparently lost for an observer outside of the BH.

Finally even Hawking seems to have changed his mind concerning
this issue and recently he has announced \cite{Hawking3} a new
mechanism  that could avoid unitarity violations in BH evaporation
(the details are yet unpublished) .

\section{Entropy Bounds}
\label{sec:3}

In 1981 Bekenstein proposed the {\it Universal Entropy Bound}
(UEB) \cite{Bekenstein2} which states that the entropy $S$ of a
complete physical system in asymptotically flat $D=4$ space-time,
whose total mass-energy is $E$, and which fits inside a sphere of
radius $R$, is necessarily bounded from above:
\begin{equation}
S \le 2\pi E R
\end{equation}
This bound can be obtained through a {\it gedanken} experiment in
which a weakly self-gravitating  object with some entropy content
is left with the lesser possible energy at a BH horizon.
Apparently the bound above follows if one wants to avoid any
violation of the GSL. The UEB is important because it is an
attempt to set limits on the entropy of system which is
characterized by physical parameters such as energy and size. This
kind of bound could be relevant, at least in principle, for
information theory, both classical and quantum (see
\cite{Galindo2} for a recent and very complete review). Since the
seminal Shannon's works the main topic of information theory has
been information transport, or in other words, channel capacity,
noise, redundancy, cryptography and many other things which have
to do with information communication. However the UEB refers to
information storage capacity. Usually this issue is contemplated
under the point of view of the smaller physical system capable to
store one single bit of information like molecules, atoms, photon
polarization etc. The point of view of the UEB is completely
different since it offers a more holistic kind of bound which
applies  to the whole memory system independently of its
microscopic structure. Note also that the bound  does not contain
(in standard units) the Newton constant at all (it refers only to
weakly self-gravitating systems but this is the case of most of
lab and astrophysical systems around us). On the other hand, from
the point of view of present applications, the UEB is far from
being of any practical interest. For example for a standard music
compact disk the UEB set a maximum of storage capacity of about
$10^{68}$ bits but present technology make possible to put {\it
only} $10^{10}$ bits on it. However, as a matter of principle, the
UEB defines a new fundamental relation between energy, size and
information.

Another kind of bound on the entropy (information) content of
non-BH objects was proposed by Susskind in 1995 \cite{Susskind1}.
According to it the maximum entropy $S_m$ of a system that can be
enclosed by a spherical surface of area $A$ is given by:
\begin{equation}
S_{m} \le \frac{A}{4}
\end{equation}
This bound is known as the {\it Spherical Entropy Bound} and it
requires the space-time to be asymptotically flat. The proper
definition of $A$ also assumes that the system has spherical
symmetry or that it is weakly gravitating. The bound is motivated
by considering another {\it gedanken} experiment called the
Susskind process. One consider a system of mass $E$ inside an area
$A$ which is smaller than the mass $M$ that would produce a BH if
fitted in the same area $A$. Now we add an infalling spherical
shell of mass $M-E$ in order to collapse the system. Then applying
the GSL to the process the bound follows.

The {\it Spherical Entropy Bound} can be derived, under some
conditions, from the UEB so in the cases where both can be applied
the former is weaker than the latter. However the {\it Spherical
Entropy Bound} is much more appropriate to introduce the main
ideas underlaying the {\it Holographic Principle}. Note that this
bound is telling us that the maximum information that a system can
store scales basically with the area of its external surface. This
is in clear contradiction with our normal experience  according to
which the information capacity scales with the volume.

The {\it Spherical Entropy Bound} can also be extended to a much
more general bound called the {\it Space-like Entropy Bound }. The
formulation of this bound is the following \cite{Bousso1}: Let be
a compact portion of equal time spacial hypersurface in space-time
with volume $V$ and boundary $B$ of area $A(B)$, then the total
entropy inside the volume $V$ is bounded by:
\begin{equation}
S(V) \le \frac{A(B)}{4}
\end{equation}
This bound seem to be the natural generalization of the {\it
Spherical Entropy Bound}. It works in many systems but it is easy
to find counter examples  where it does not apply, meanly in
cosmological scenarios or for strongly gravitational systems (see
\cite{Bousso1} for a detailed discussion). The failure of this
bound lead Bousso to propose a suitable generalization of this
bound which is known as the {\it Covariant Entropy Bound} that
will be discussed latter.

\section{The 't Hooft Holographic Principle}
\label{sec:4} After almost one decade of lonely efforts outside of
the main stream in theoretical physics 't Hooft \cite{tHooft}
(followed by Susskind \cite{Susskind1}) presented his {\it
Holographic Principle}. This principle changes radically our
thinking about the counting of degrees of freedom of physical
systems and the way in which the entropy or information content is
stored.

There is no any well established enunciate of this principle but
according to the ideas of 't Hooft and Susskind one possible
preliminary formulation of the {\it Holographic Principle} could
be:

 The full physical description some given region $R$, in an $D$
dimensional universe, with $D-1$ dimensional boundary $B=\partial
R$, can be reflected in processes taken place in $B$.

Clearly the above formulation is too vague to be of practical
interest but still the {\it Holographic Principle} is regarded as
a major clue in the search for the solution for the Quantum
Mechanics versus GR conundrum. In particular any fundamental
theory should incorporate this counterintuitive result.

In fact we have already an example where the {\it Holographic
Principle} could be taking place, namely the Maldacena conjecture
known as the {\it AdS/CFT correspondence} \cite{MW}.  According to
it, for some given settings, the physics of a string theory of
type IIB, defined on an $AdS_5 \times S^5$ space, is equivalent to
the physics of a maximally supersymmetric Super Yang-Mills $U(N)$
theory defined on the boundary of the $AdS_5$ space. Even in the
absent of a real proof, the Maldacena conjecture has passed
successfully a great number of checks and it is generally believed
to be true.

In spite of the fact that the {\it Holographic Principle} is not
yet defined in a precise way, it is clear that, from a fundamental
point of view, the entropy bounds discussed in the previous
section are likely to be a more or less straightforward
consequence on this principle. In any case, in the absence of a
well formulated {\it Holographic Principle}, entropy bounds  can
be extremely useful as heuristic tools for the task of clarify the
apparent contradictions between Quantum Mechanics and GR.

\section{The Covariant Entropy Bound}

In order to solve the difficulties found in the {\it Space-like
Entropy Bound}, Bousso proposed the {\it Covariant Entropy Bound}
(CEB) \cite{Bousso2,Bousso1} . The bound can be stated as follows:
Let $B$ be any spatial $D-2$ dimensional hypersurface with area
(volume) $A(B)$. A $D-1$ dimensional hypersurface $L$ is called a
light-sheet of $B$ if it is generated by a congruence of null
geodesics beginning at $B$, extend orthogonally from $B$ and has
negative expansion. Now we define $S$ as the entropy of any matter
${\it illuminated}$ by the $B$'s light sheet. Then
\begin{equation}
S(L) \le \frac{A(B)}{4}
\end{equation}
In order to clarify the meaning of the CEB note that for any point
in $B$ it is possible to construct four light rays (branches). Two
of them go to the future and two go to the past. On any of these
branches, a ray, together with its neighbors, defines a positive
or a  negative expansion (rays converging or diverging). The $L$
set considered in the above formulation of the CEB is the one
corresponding to future going converging congruence. The rays so
defined may end at the tip of a cone (for spherical symmetry) or
more generally on a caustic. After this the rays will start to
diverge but this region is considered to be outside of $L$. The
entropy of the matter traversed by this set of rays before they
reach the caustic is the one bounded according to the Bousso's
covariant entropy bound.

The CEB can successfully solve the difficulties found by the {\it
Space-like Entropy Bound}, in particular in cosmological and
strong gravitating scenarios. In fact it  holds in a wide range of
situations and no physically realistic counter example has been
found so far. A logical consequence of this is to consider the CEB
as a hint for more fundamental law of nature. According to Bousso
the {\it Covariant Holographic Principle} could be stated (using
the previous notation) as follows: The fundamental theory
underlying Quantum Mechanics and GR should be such that the matter
and the geometry illuminated by the convergent rays starting from
$B$ have a number of independent states ${\cal N}(L(B))$ which is
bounded by:
\begin{equation}
{\cal N}(L(B))  \le   e^{A(B)/4}
\end{equation}
If the fundamental theory contains Quantum Mechanics in its
present form, the number of (quantum) states ${\cal N}$ is just
the dimension of the Hilbert Space but this does not have to be
necessarily the case. In the language  of information theory  the
number of bits times $ln 2$ involved in the description of $L(B)$
must be bounded by $A(B)/4$.

There is also a stronger version of the CEB which was proposed in
\cite{FMW} known as the {\it Generalized Covariant Entropy Bound}
(GCEB). In this version the light rays in $L$ are allowed to stop
before they reach the caustic and in this way they define a new
surface $B'$ of area $A(B')$. Then according to the GCEB we have:
\begin{equation}
S(L) \le \frac{A(B)-A(B')}{4}
\end{equation}
Obviously the GCEB reduces to the CEB for the particular case
$A(B')=0$. In \cite{FMW} and \cite{Bousso3} some different sets of
necessary conditions for the energy and entropy matter content
are shown to guarantee the validity of the GCEB and it has being
shown that the UEB can be obtained from the GCEB \cite{Bousso4}.
It has also been argued that CEB  must be modified in some way in
order to include Hawking radiation \cite{Strominger1}.

\section{Discussion}

Independently of the precise formulation of the {\it Holographic
Principle} or its possible consequences, for example  the entropy
bounds, it is apparent that there are many indications that point
towards the validity of some law of a similar kind. Therefore this
hypothetical law  should be encoded in any fundamental theory that
may reconcile  Quantum Mechanics and GR. However, even in the
absence of such a theory, there are some important conclusions
that can be drawn from the {\it Holographic Principle}.

The first one is that Quantum Field Theory (QFT) does not work
when strong gravitational effects are present. In order to see why
this is the case we can consider a QFT defined in a finite volume
$V$ (to avoid infrared divergencies). Then the divergencies are
ultraviolet and typically they are regulated by means of an energy
cutoff  (or equivalently by introducing some minimal distance ).
In any case the number of degrees of freedom scales with the
volume $V$  of the system and not with the external area as the
{\it Holographic Principle} seems to suggest. The reason for this
is that even when the theory is   regularized by cutting the high
energy modes, there remain an enormous number of field
configurations that are gravitationally unstable and would
collapse into a BH. By removing all of these configurations we are
left with a much less number of states that would scale with the
external area of the system in a way compatible with the {\it
Holographic Principle}.

On the other hand many people consider Superstring Theory as the
most sound candidate for a quantum theory of gravity. It has been
argued that Superstring Theory violates the {\it Holographic
Principle} since the number of states scales also with the volume
of the system. However this is true only at the perturbative
level. When non-perturbative effects are taken into account  in
Superstring Theory one has to deal with the so called M-theory.
The non-perturbative regime can be reached in some cases by means
of some duality transformation from the perturbative one but the
physics turn out to be completely different in general. In
particular strings give rise to new bound states  (D-branes) in
such a way that the counting of degrees of freedom may change
drastically from the perturbative regime. Moreover it has been
shown that the perturbation series breaks down before the
holographic bounds are reached.  In fact the {\it AdS/CFT
Correspondence} and the successful computation of the entropy in
some kind of BH, done in the framework of M-Theory by counting
microscopic states \cite{Strominger2}, seems to suggest that
M-theory could be compatible with the {\it Holographic Principle}.

In any case one can consider the {\it Holographic Principle} as a
necessary ingredient of any fundamental theory and use it in order
to make predictions even if the fundamental theory is far from
being established. Finally, as in any other branch of physics,
confrontation with the experiment will be the final test for this
principle and if the general idea is correct it can play an
important role in the future.

To end  it is important to stress that in this paper we have not
considered the possibility of having a non-zero cosmological
constant. However recent data, coming from the  cosmic microwave
background radiation, distant supernovae, and the spectrum of the
density fluctuations, strongly suggest that the cosmological
constant is different from zero. If this is the case almost all
the points mentioned in this article should be reconsidered even
if for many applications the cosmological constant can be safely
neglected. Work is in progress in this direction.

\section{Acknowledgments}

This work has been partially supported by the DGICYT (Spain) under
the project numbers numbers FPA 2000-0956 and BFM 2002-01003. The
author acknowledges  the hospitality of the SLAC Theory Group,
where the final part of this work was done, economical support
from the Universidad Complutense del Amo Program and congratulates
Alberto Galindo for their 70th birthday.

%
%


\begin{thebibliography}{99.}
%
%
%


\bibitem{Einstein1} Einstein, A, (1905)
Annalen der Physik {\bf 17}: 132, 549, ,639, 891


\bibitem{Einstein2} Einstein, A, (1915)
Sitzungsber Preuss. Akad. Wiss. (Math. Phys.),  Berlin: 844

\bibitem{Michell} Michell, J, (1784)
Phil. Trans. R. Soc. {\bf 74}:35

\bibitem{Laplace} Laplace, P S, (1796)
Le Systeme du Monde, Vol.II Paris: 305


\bibitem{KN} Newman, E T et al, (1965)
J. Math. Phys. {\bf 6}:918

\bibitem{MTW} Misner, C W, Thorne, K S and Wheeler, J A, (1973)
Gravitation, Freeman, New York

\bibitem{HandE} Hawking, S W  and Ellis, G F R, (1973)
The large scale structure of space-time, Cambridge University
Press, Cambridge

\bibitem{Wald} Wald, R M, (1984)
General Relativity, The University of Chicago Press, Chicago


\bibitem{BHproperties} Penrose, R and Floyd, R M, (1971) Nature {\bf 229}:177\\
Christodoulou, D, (1970) Phys. Rev. Lett. {\bf  25}:1596\\
Hawking, S W, (1971) Phys. Rev. Lett. {\bf  26}:1344

\bibitem{Galindo1} Galindo, A and Mas,  L, (1981)
Soluciones Exactas en Relatividad General. Colapso Gravitacional y
Agujeros Negros, Editorial de la Universidad Complutense, Madrid:
163



\bibitem{Bekenstein1} Bekenstein, J D, (1972) Nuovo Cim. Lett. {\bf 4}:737   \\
(1973) Phys. Rev. {\bf D7}:2333  \\
(1973) Phys. Rev. {\bf D9}:3292

\bibitem{Hawking1} Hawking, S W, (1974)
Nature {\bf 248}: 30  \\
(1975) Commun. Math. Phys. {\bf 43}: 199

\bibitem{Hawking2} Hawking, S W, (1976)
Phys. Rev. {\bf D14}: 2460


\bibitem{Hawking3} Hawking, S W, (2004)
Talk given in the General Relativity 17th Dublin Conference

\bibitem{Bekenstein2} Bekenstein, J D, (1981)
Phys. Rev. {\bf D 23}:287


\bibitem{Galindo2} Galindo, A and  Mart\'\i n-Delgado, M A, (2002)
 Rev. Mod. Phys. {\bf 74}:347



\bibitem{Susskind1} Susskind, L, (1995)
J. Math. Phys. {\bf 36}:6377


\bibitem{Bousso1} Bousso, R, (2002)
Rev. Mod. Phys. {\bf 74}:825



\bibitem{tHooft} 't Hooft, G, (1993)
Salam-festschrifft, World Scientific, Singapour\\
 gr-qc/9310026

\bibitem{MW} Maldacena, J, (1998)
Adv. Theor. Math. Phys. {\bf 2}:231  \\
Witten, E, (1998) Adv. Theor. Math. Phys. {\bf 2}:253

\bibitem{Bousso2} Bousso, R, (1999)
JHEP {\bf 9907}:004

\bibitem{FMW} Flanagan, E E, Marolf, D and Wald, R M, (2000) Phys. Rev. {\bf
D62}:084035

\bibitem{Bousso3} Bousso, R, Flanagan E E and Marolf, D, (2003)
Phys.Rev. {\bf D68}:064001

\bibitem{Bousso4} Bousso, R, (2004)
JHEP {\bf 0405}:050

\bibitem{Strominger1} Strominger, A and Thompson, D, (2004)
Phys.Rev. {\bf D70}:044007


\bibitem{Strominger2} Strominger, A and Vafa, C, (1996)
Phys. Lett. {\bf B379}:99





\end{thebibliography}
\end{document}